\documentclass[aps,prl,twocolumn,showpacs,groupedaddress]{revtex4}
\usepackage{graphicx}  % needed for figures
\usepackage{dcolumn}   % needed for some tables
\usepackage{bm}        % for math
\usepackage[english]{babel}
\usepackage{amsfonts,amsmath,amssymb,mathrsfs}
\usepackage{epstopdf}
\usepackage{subfigure}
\usepackage{color}

\def\a{\alpha}
\def\r{\rho}
\def\s{\sigma}
\def\t{\tau}
\def\m{\mu}
\def\n{\nu}
\def\k{\kappa}
\def\th{\theta}
\def\g{\gamma}\def\G{\Gamma}
\def\L{\Lambda}\def\l{\lambda}
\def\D{\Delta}
\def\la{\langle}
\def\ra{\rangle}
\def\o{\omega}\def\O{\Omega}
\def\d{\delta}
\def\p{\partial}

\def\half{\textstyle{\frac{1}{2}}}

\def\bdoc{\begin{document}}
\def\edoc{\end{document}}

\def\beq{\begin{equation}}
\def\eeq{\end{equation}}
\def\bea{\begin{eqnarray}}
\def\eea{\end{eqnarray}}
\def\ben{\begin{enumerate}}
\def\een{\end{enumerate}}
\def\la{\langle}
\def\ra{\rangle}
\def\a{\alpha}
\def\b{\beta}
\def\g{\gamma}\def\G{\Gamma}
\def\d{\delta}\def\D{\Delta}
\def\e{\epsilon}

\def\th{\theta}
\def\k{\kappa}
\def\l{\lambda}
\def\m{\mu}
\def\n{\nu}
\def\o{\omega}
\def\p{\pi}
\def\r{\rho}
\def\s{\sigma}
\def\t{\tau}
\def\L{{\cal L}}
\def\S{\Sigma }
\def\gsim{\; \raisebox{-.8ex}{$\stackrel{\textstyle >}{\sim}$}\;}
\def\lsim{\; \raisebox{-.8ex}{$\stackrel{\textstyle <}{\sim}$}\;}
\def\gtrsim{\gsim}
\def\lessim{\lsim}
\def\loc{{\rm local}}
\def\vm{v_{\rm max}}
\def\bh{\bar{h}}
\def\del{\partial}
\def\nab{\nabla}
\def\half{{\textstyle{\frac{1}{2}}}}
\def\fourth{{\textstyle{\frac{1}{4}}}}

\def\bD{{\bf D}}
\def\bE{{\bf E}}
\def\bF{{\bf F}}
\def\bB{{\bf B}}
\def\bP{{\bf P}}
\def\bV{{\bf v}}
\def\bv{{\bf v}}
\def\bx{{\bf x}}
\def\by{{\bf y}}
\def\bz{{\bf z}}
\def\ba{{\bf a}}
\def\bd{{\bf d}}
\def\bs{{\bf s}}
\def\bn{{\bf n}}
\def\bp{{\bf p}}

\def\O{\Omega}

\def\br{{\bf r}}
\def\bnab{{\bf \nab}}

\def\tE{\tilde{E}}
\def\tL{\tilde{L}}

\begin{document}

\title{Over-spinning a black hole with a test body}
\author{Ted Jacobson$^*$ and Thomas P. Sotiriou$^{*\dagger}$}
\affiliation{$^*$Center for Fundamental Physics,  University of Maryland, College Park, MD 20742-4111, USA}
\affiliation{$^\dagger$Department of Applied Mathematics and Theoretical Physics, Centre for  
Mathematical Sciences, University of Cambridge, Wilberforce Road,  
Cambridge, CB3 0WA, UK}
\date{\today} 
\begin{abstract}
It has long been known that a maximally spinning black hole can not be 
over-spun by tossing in a test body. Here we show that if instead 
the black hole starts out with below maximal spin, then indeed 
over-spinning can be achieved
when adding either orbital or spin angular momentum. We find that requirements on the 
size and internal structure of the test body can be met as well. 
Our analysis neglects radiative and self-force effects,
which may prevent the over-spinning.

\end{abstract}  
\pacs{04.70.Bw Classical black holes, 04.20.Dw	Singularities and cosmic censorship}
\maketitle

%\section{Introduction}

Gravitational collapse in general relativity 
inevitably leads to spacetime singularities~\cite{Penrose:1964wq,Hawking:1969sw}
where the theory presumably breaks down.
According to the ``Cosmic Censorship" conjecture~\cite{Penrose:1969pc}, 
``naked singularities" --- i.\ e.\ singularities visible from afar --- cannot be 
formed via any process that involves physically reasonable matter.
Although this hypothesis remains unproven, 
significant evidence suggests that it may indeed be true, in the
sense that whenever singularities arise from generic, nonsingular initial
data, they may be always hidden behind a black hole event horizon
(and may even be invisible to {\it any} observer) \cite{Wald:1997wa,Penrose:1999vj}.
One way of testing Cosmic Censorship is thus 
to ask whether it may be possible to destroy a black hole horizon.
 
The general stationary, vacuum black hole solution in 
general relativity is a Kerr black hole, which is characterized 
by its mass $M$ and angular momentum $J$.
For angular momentum greater than the ``extremal" limit
$GM^2/c$,
the Kerr metric has no horizon and has a
naked singularity.
The question we are posing is
whether an object with either orbital or spin angular 
momentum---hereafter called 
the ``body"---can be
dropped into a Kerr black hole
so as to push the resulting composite object 
over the extremal limit. If so, that object could
not settle down to a stationary black hole. 
Presumably it would form a naked singularity,
in violation of Cosmic Censorship.

This question was
considered and answered in the negative 
long ago by Wald~\cite{Wald} under the 
assumptions that the black hole is initially extremal and the body 
can be treated as a test particle.  
Here we relax the first assumption, considering an initial black 
hole that is nearly but not quite extremal, and examining whether one 
can ``jump over" the extremal limit
in the test body approximation. 
Our question is motivated by Hubeny's analysis~\cite{Hubeny:1998ga}
which showed in an analogous setting that a charged test particle
can be added to a charged black hole in such a way that the
resulting object has too much charge to be a black hole,
and by the analysis
of Hod~\cite{Hod:2002pm}
which showed that a point test particle can be
dropped from the horizon into a spinning black hole in
such a way that the resulting object has too much angular momentum
to be a black hole. 

We extend Hod's result, 
finding a broader class of trajectories by which over-spinning
may be accomplished, and we also consider physical 
structure and size limitations of the particle. 
It turns out that the latter do not 
preclude the over-spinning.
Radiation and/or
self-force effects, which lie outside the 
test body approximation, might nevertheless 
prevent the over-spinning.
We note that the considerations of the
present paper are strictly classical, in contrast to
a number of recent studies of quantum 
mechanical tunneling processes that might be able to over-spin
a black hole (see \cite{Matsas:2009ww}, and references
therein). Hereafter we adopt units with $G=c=1$.

In order to be able to treat the body as a 
test body, we shall assume its energy $\d E$ and 
angular momentum $\d J$, defined with respect to the
Killing vectors of the black hole, are 
small compared to those of the black hole, 
\beq\label{small}
\d E\ll M,\qquad \d J\ll J.
\eeq
Note that the black hole must start out 
very close to extremal if the small perturbation 
caused by the body is to have any chance of 
pushing it over the extremal limit.

%\section{Bounds on the energy and angular momentum added}

%\subsection{over-spin condition}

We assume that the angular momentum of the body is aligned 
with the spin of the black hole, in order to maximize the angular momentum added. 
The question, then, is whether $\d E$ and $\d J$
can be chosen such that the body falls
into the black hole, with negligible corrections to the
test body approximation, and such that 
\beq
J+\d J > (M+\d E)^2.
\eeq
This sets the lower bound on the required 
angular momentum carried by the body,  for a 
given $\d E$:
\beq\label{over-spin}
\d J > \d J_{min}=(M^2-J)+2M \d E+\d E^2.
\eeq
Since we are assuming $\d E\ll M$, it might seem
that the $\d E^2$ term may as well just be neglected
at this stage, but in fact that term imposes 
an upper bound on $\d E$ and $\d J$. 

Equation (\ref{over-spin}) yields
\beq\label{largespin}
\d J/\d E^2>2M/\d E \gg1.
\eeq
If $\d E$ were equal to the rest mass of the
body, 
and if $\d J$ comes from spin,
this would imply that the body
has angular momentum far over the 
extremal ratio. In itself this is not a problem,
since there is no a priori upper limit to this ratio
if one is not restricting to black holes.
Note also that $\d E$ can be much less than the 
rest mass, if the body is deeply bound.

%\subsection{Fall in condition}
An upper bound on the angular momentum of the body comes from the 
requirement that it does indeed cross the horizon to end up in the black hole.
Wald~\cite{Wald} derived such a bound by analyzing the trajectories using the geodesic or Papapetrou equations for the case of orbital angular momentum and spin respectively. We will use another path to the same results which does not actually require the geodesic or Papapetrou equations.

If the body falls across the horizon, then the flux of 
energy and angular momentum into the black hole
are related to the stress energy tensor of the body via
\beq
\d E-\O_H\d J=\int T_{ab}\chi^a d\Sigma^b.
\eeq
Here $\O_H=a/2Mr_+$ is the angular velocity of the horizon,
where $a=J/M$ the specific 
angular momentum and 
$r_+=M+(M^2-a^2)^{1/2}$ is the horizon radius
in Boyer-Lindquist 
coordinates. The vector
 $\chi^a=\del_t^a+\O_H\del_\phi^a$ is the horizon-generating Killing
 vector, and $d\Sigma^b$ is the horizon surface element.
 Both $\chi^a$ and $d\S^b$ are parallel to the null generator 
 of the horizon, so the null energy condition on the 
 matter of which the body is composed implies
 \beq\label{go-in2}
\d E>\O_H\d J,
\eeq
 which can be written as
 \beq\label{go-in}
\d J < \d J_{max}=\frac{2Mr_+}{a} \d E.
\eeq
This constraint guarantees that the body can 
fall across the horizon starting from {\it some} point outside,
although in general it is in a bound orbit that does not come 
from spatial infinity.

%\subsection{over-spinning process}

As long as $\d J_{min}<\d J_{max}$ for some
$\d E$, there will 
be values of $\d J$ satisfying both inequalities
(\ref{over-spin}) and (\ref{go-in}).
If the black hole starts out extremal ($J=M^2$, 
so $a=M=r_+$),   
then $\d J_{min}=2M\d E +\d E^2$ is {\it never} less than
$\d J_{max}=2M\d E$, so it is impossible to over-spin
the black hole. This was observed long ago by 
Wald~\cite{Wald}. In the sub-extremal case, however,
these inequalities {\it can} be satisfied.

To understand the nature of the allowed
range it is helpful to visualize the inequalities
graphically. If $\d J_{max}$ and $\d J_{min}$ 
are plotted vs. 
$\d E$, 
the former is a straight line through the
origin, with slope $2M r_+/a>2M$,
while the latter is a parabola with positive
intercept, slope $2M$ at the intercept, and 
curved upwards. Some algebra reveals that 
the parabola always intersects
the straight line in two points. The allowed
values of $\d E$ and $\d J$ are those in the
compact region above the parabola and below the
straight line. Note that
if the $\d E^2$
is neglected in (\ref{over-spin}),
the parabola is replaced by a straight line,
and no upper bound is imposed on the 
allowed values. 
The case considered by Hod~\cite{Hod:2002pm},
i.e.\ that of dropping the particle from a point on the horizon,
corresponds to the upper boundary of this region, $\d J =\d J_{max}$.

Rather than giving exact formulae,
it is more illuminating to 
expand in the small dimensionless quantity
$\e\ll1$ defined by 
\beq\label{epsilon}
J/M^2=a/M=1-2\e^2.
\eeq
(Hubeny~\cite{Hubeny:1998ga} used the same parameter to analyze the 
charged case.)
Also at this stage we adopt units with $M=1$, 
which will keep the expressions simpler. 
Then we have 
\bea\label{minmax}
\d J_{min} &=& 2\e^2+2\d E + \d E^2\\
\d J_{max}&=& (2+4\e)\d E,\label{jmax}
\eea
where terms of order $O(\e^2\d E)$ have been dropped
in (\ref{jmax}).
The allowed range of $\d E$ lies where the difference
\beq\label{max-min}
\d J_{max}-\d J_{min} = 4\e\d E - \d E^2 -2\e^2
\eeq
is positive, i.\ e.\
\beq\label{dEe}
(2-\sqrt{2})\e <\d E<(2+\sqrt{2})\e.
\eeq
In particular, $\d E$ must be of order $\e$,
which is consistent with the requirement
(\ref{small}) that the body make only a small perturbation. 
For a given $\d E$, the allowed values of 
$\d J$ are near $2\d E$, so we must have 
\beq\label{dJ/dE}
\d J\sim \d E.
\eeq
Note that the {\it width} (\ref{max-min}) of the allowed
range of $\d J$ is only of order $\e^2\ll\e$.
%The midpoint of the allowed region for $\d E$ 
%is $\d E = 2\e$, for which the allowed range of 
%$\d J$ is $4\e+6\e^2<\d J < 4\e + 8 \e^2$. 

As already mentioned, 
the black hole must start out
very nearly extremal, but now we can be
somewhat more quantitative. According to (\ref{small})
and (\ref{dEe}) we must have $\e\ll1$,
and $a - 1=2\e^2$ is parametrically smaller.
For example, if $\e=10^{-2}$, then 
the initial black hole must have 
$a=0.9998$. For a thought experiment
we can imagine even smaller values of $\e$. 
We conclude that, if the body can be treated as a
point test particle, the black hole
can indeed be over-spun.

We turn now to consideration of the finite size
requirements for the body,  beginning with the case
of orbital angular momentum in the equatorial 
plane. Here the issue
is that in order to have the required values of 
$\d E$ and $\d J$, the body might have to be in a bound orbit,
which would have a turning point at a maximum radius.
In that case we would need to require that
the body be small enough to fit outside the horizon at this radius.
Since the body can be no smaller than a black hole
with the same rest mass, it is not clear in advance whether
this requirement could be met. However, we find that
this size constraint is not an issue, since in fact there are 
orbits that come in from infinity with no turning point.

To address this point we recall that the proper time derivative
of the (Boyer-Lindquist) radial coordinate of orbital motion in the equatorial 
plane satisfies $\dot{r}^2/2 + V_{\rm eff}(r,\tE,\tL)=0$, 
where the effective potential is given in terms of the 
energy $\tE$ and angular momentum $\tL$ per unit mass by~\cite{Wald:1984rg}
\beq
V_{\rm eff}= -\frac{1}{r}+\frac{\tL^2}{2r^2}
-\frac{(\tL-a\tE)^2}{r^3}
+ \frac{1}{2}(1-\tE^2)\left(1+\frac{a^2}{r^2}\right).
\eeq
The turning points are located where $V_{\rm eff}(r)=0$. 
For the energy $\d E$ we 
choose the value at the center of the 
allowed region given above, so 
$\tE = 2\e/m$, where $m$ is the rest mass of the 
body. The allowed range of the specific angular momentum
is then $ (2+3\e)\tE<\tL<(2+4\e)\tE$, so we are led to
parametrize the specific angular momentum as
$\tL=(2+b\e)\tE$, where $3<b<4$. 
If values of $\e$, $m$ and $b$ can be found
for which $V_{\rm eff}<0$ everywhere outside the horizon,
then with such values a body can fall all the way 
into the black hole from infinity.

We explored this question numerically and found that
such orbits indeed exist. Two examples with $\e=0.01$
and $b=3.3$ 
are $m=0.01$ ($\tE=2$) and $m=0.001$
($\tE=20$).
To understand a bit more  quantitatively we can
focus on large asymptotic velocity $\tE\gg1$, dropping 
$\tE$ independent terms and expanding the potential
out to second order in $\e$, which yields
\bea
V_{\rm eff} = 
-\frac{\tE^2}{2}\Bigl[1 &-&\Bigl(3 + 4b\e + (b^2 + 4)\e^2\Bigr) r^{-2}
\nonumber\\
&+& \Bigl(2 + 4b\e +2(b^2+4)\e^2\Bigr)r^{-3}\Bigr].
\eea
This potential is negative at $r=0$ and at $r=\infty$,
and has a single maximum at $r=1 + 2b\e/3+O(\e^2)$,
where it is equal to $-(2-b^2/6)\tE^2\e^2 +O(\e^3)$. 
Hence it is everywhere negative provided $b<\sqrt{12}\simeq 3.46$,
in which case the body falls across the horizon.

%\section{Size considerations}

Next we examine whether requirements relating to the 
size and structure of the body can be met in the spinning case.
For simplicity we assume that the body 
is dropped along the symmetry axis.
%\subsection{Weakly bound case}
We first consider the case when $\d E\sim m$,
and the body is not spinning relativistically, so its spin
angular momentum
is given by $\d J\sim mvR\sim \d E\, v R$, where 
$v$ is the surface velocity and 
$R$ is the equatorial radius. The condition $v<1$ then implies
$R>\d J/\d E$. We saw above that
the ratio $\d J/\d E$ must be of order unity (\ref{dJ/dE}), 
that is
of order $M$. In this case the body must be larger than
the black hole, so it simply will not fit in the transverse
direction, and 
in any case treating it as a point particle with spin
would be unjustified,
since that rests on the
smallness of the size of the body compared to the 
ambient radius of curvature. Moreover, the radial tidal 
stress required to hold the body together 
would be larger than the energy density, violating
energy conditions.

It cannot help to allow ultra-relativistic tangential
velocity. The reason can be seen with a simple
Newtonian estimate. If the 
surface acceleration $v^2/R$ 
exceeds $\sim 1/R$, the required
force exceeds $m/R$. But what could provide this
force to hold the body together?
The self-gravitational force $\sim m^2/R^2$ 
cannot produce this acceleration unless $m/R >1$,
which is the condition that the body becomes
a black hole. A black hole cannot 
satisfy (\ref{largespin}), so this will not do. 
Alternatively, suppose the force is provided
by internal tension. Then the $T_{\hat{r}\hat{r}}$ 
stress tensor component in an orthonormal rest frame
must satisfy $-T_{\hat{r}\hat{r}}R^2> m/R$, 
hence $-T_{\hat{r}\hat{r}}> m/R^3$. But the
right hand side is the
energy density of the
body,  so this inequality violates all the energy 
conditions. The body would therefore need to be 
composed of unphysical material. 
The conclusion is that it is impossible to 
over-spin the black hole if the body's Killing energy is 
close to its rest mass, $\d E\sim m$.

%\subsection{Deeply bound case}

Since the angular momentum involves the
rest mass $m$, not the Killing energy,
it might be possible to achieve a large enough 
$\d J$ with a small enough size $R$, 
without requiring unphysical matter,
by dropping the body from a position
where it is deeply bound, $\d E\ll m$. 
This might be achieved by slowly lowering the
body on a tether,  down to the near the
black hole horizon, before dropping it in.
Now we reconsider whether the size restrictions 
can be met in this setting. 

We begin with the restrictions on the rest mass $m$. 
If $m$ is much greater than $\d E$, then
the test body approximation requires
that we impose not only $\d E\ll 1\, (=M)$ (\ref{small}),
but also $m\ll 1$. There is also a lower bound on $m$,
coming from an upper bound on $R$: 
the angular momentum is  
$\d J\sim mvR$,
hence
(restricting to nonrelativistic spin $v<1$ as required
by the previous analysis)
$R > \d J/m\simeq 4\e/m$.
The requirement $R\ll1$ then yields
$m\gg\e$. The mass and size must therefore fall within
the ranges
\beq\label{bounds}
\e\ll m\ll 1, \qquad 4\e/m\lessim R\ll 1.
\eeq

The inequality (\ref{go-in2})
guarantees that the body can cross the horizon with the chosen values
of energy and angular momentum, but since the deeply
bound drop point lies at a finite distance from the horizon it 
is necessary to check that (a) the spinning body would actually fall into the
black hole rather than being repelled, and (b) the body polar radius, $R_{\rm polar}$, 
can be chosen
smaller than
the proper distance $d$ from the horizon to the drop point
\beq\label{R<d}
R_{\rm polar}<d,
\eeq
so that the body can 
fully fit outside the black hole and be localized at the drop point.   

To address these issues, we note~\cite{Mino:1995fm}
that a spinning body of rest mass $m$,
moving along the axis of rotation, 
has energy $\d E$ and angular momentum $\d J$ 
satisfying
\beq\label{conserved}
\d E  - \O(r)\d J =m\gamma(r)  N(r)
\eeq
where 
$\g(r)$ is the gamma factor for the body relative to the 
rest frame of the timelike Killing vector, and 
\beq
N(r)=\sqrt{\frac{r^2-2r+a^2}{r^2+a^2}}, \qquad
\O(r)=\frac{2ar}{(r^2+a^2)^2}
\eeq
respectively are the norm of the Killing vector 
(which vanishes at the horizon)
and the limiting angular velocity of ZAMOs
(zero angular momentum
stationary observers) as 
the axis of rotation is approached.
The body can exist at rest at 
radial coordinate $r$, with the given values
of the constants of motion, if
(\ref{conserved}) is satisfied with $\g(r)=1$. 
The body will
fall in across the horizon from
such a point provided (i) 
the left hand side
of (\ref{conserved}) is positive
at the horizon, i.e.\
(\ref{go-in2}) holds, and (ii)
$r$ is the closest radius to the
horizon where (\ref{conserved}) is satisfied
with $\gamma(r)=1$.
%Both sides of (\ref{rest})
%are monotonic increasing from the 
%horizon.
%The left hand side starts out positive and
%changes slowly,
%the right hand side starts out at zero,
%and the drop point we are interested in corresponds to
%the first intersection of their graphs. 
%As $m$ grows, that intersection
%moves inward, i.e.\ to smaller radii.

To estimate the maximum distance 
from the deeply bound 
drop point to the horizon, for a given 
mass $m$, we make use of the fact that 
$N\ll1$ there.
In this regime, and for a near-extremal black hole, 
$\O(r)\simeq \half -\sqrt{\e^2+N^2/2}$,  and 
so for the required values of energy and
angular momentum we have $\d E  - \O(r)\d J = O(\e^2) + O(\e N)$. 
This must be equal to $mN$, which according to
(\ref{bounds}) must be $\gg\e N$.
It follows that $\e\gg N$, hence the horizon
value $\O_H\simeq \half -\e$ may be used for $\O(r)$.
Using $\d J_{min}$ (\ref{minmax}) 
we find  $(\d E  - \O_H\d J)_{max} = \e^2$,
(which occurs for $\d E=2\e$).
On the other hand, near the horizon we 
have
$N\simeq \k d\simeq \e d$,
where $\k$ is the surface gravity (equal
to $\lim_{r\rightarrow r_+}|\nab N|$) and $d$ is
the proper distance to the horizon
(orthogonal to the surfaces $N=$ const).
Hence the maximum value $d$ can take is
$d\simeq \e/m$.
According to (\ref{bounds}) we must therefore have $d\ll1$,
which justifies having used of the linear
near-horizon approximation to $N$~\footnote{More generally, 
in the 
near-horizon region ($N\ll1$) of a near-extremal
black hole ($\e\ll1$)
one finds
$N(r)\simeq \sqrt{2}\e \sinh (d/\sqrt{2})$.}.
To fit outside the horizon at the drop point, 
the body must therefore satisfy 
\beq
R_{\rm polar}<\e/m.
\eeq
This bound and the condition (\ref{bounds}) on the
equatorial radius $R$ can be simultaneously
satisfied, provided the body is at least 
somewhat oblate, with $R>4R_{\rm polar}$. 
(For an earlier appeal to oblate bodies in this context see \cite{deFelice:2001wj}).)

To summarize, we have considered gedanken experiments where an
object with spin or orbital angular momentum is dropped into a near-extremal 
Kerr black hole, in an effort to drive the black hole beyond the extremal limit. 
We have found that, to the extent that radiative and self-force effects can be 
neglected, the black hole
can be over-spun. In the orbital case this can be achieved even with an object dropped from infinity.
In the spin case,  the requirement that
the internal stresses satisfy the energy conditions implies that 
the object has to be deeply bound and somewhat oblate.

We saw that, from purely kinematic considerations,
the relation between the energy and angular momentum of the 
dropped body must be very finely tuned: they are both of order $\e$ in
magnitude, but the allowed window for angular momentum, given the 
energy, is only of order $\e^2$. Since the over-spinning process we
found involves a delicate balance, it is certainly possible that, 
although small, gravitational radiation and/or 
self-force~\footnote{Hod~\cite{Hod:2002pm} suggested that
the effect of self-force can be estimated by 
making the replacement $\O_H\rightarrow\O_H+\o \d J$ 
in the bound (\ref{go-in2}), and noted that if $\o>1/8$ this would
suffice to preclude over-spinning.}
effects may always manage to preclude the over-spinning. Indeed, given the 
existing evidence for Cosmic Censorship, this seems likely. 
 Our analysis suggests a dynamical regime in which
it may be interesting to study these effects.

As a final remark, we note that if over-spinning
can be achieved, then the likely formation of 
a naked singularity could  
provide an escape from Hawking's proof of
the second law of black hole mechanics~\cite{Hawking:1971tu}, that the black hole 
horizon area cannot decrease (or disappear), as well as from Israel's 
proof of the third law~\cite{Israel}, that the surface gravity cannot be reduced 
to zero in a finite time.

We thank E. Barausse, A. Buonanno, S. Liberati, and E. Poisson
for helpful discussions. This research was supported in part by
the NSF
%National Science Foundation 
under 
grant  PHY0601800,
and  T. P. S.\ was also supported by STFC. 

{\bf Note:}  The present version of this paper incorporates
the oblate case correction which appears with the published version 
only as an erratum.

%
%\section{Erratum}

%We assumed in Eqn. (17) that the body is spherical and, 
%therefore,  its polar and equatorial radii are equal.
%There is no reason to make this assumption. If we relax it, 
%then (17) should be replaced by $R_{\rm polar}<d$, and (21) 
%becomes $R_{\rm polar}<\e/m$. This eliminates the contradiction
%with (16), $R>4\e /m$, which applies to the equatorial radius. 
%The conclusion is therefore changed: It is indeed possible, 
%within the test body approximation, to overspin the black hole by 
%lowering a spinning test body to near the horizon along the 
%black hole spin axis, and dropping it in,  provided that the body 
%is somewhat oblate. Our oversight was brought to our attention 
%by reading Ref.~\cite{deFelice:2001wj}, in which the process
%of lowering a spinning test body into an extremal charged 
%black hole was analyzed, allowing for an oblate body. 

%

 \edoc